\documentclass[sigconf, nonacm]{acmart} 

\AtBeginDocument{%
  }

\usepackage{graphicx}
\usepackage{booktabs}
\usepackage{tabularx}
\usepackage{xcolor}

\usepackage{enumitem}
\setlist[itemize]{leftmargin=1.2em, labelsep=0.5em, topsep=3pt, itemsep=1pt, parsep=0pt}
\setlist[enumerate]{leftmargin=1.7em, labelsep=0.5em, topsep=3pt, itemsep=1pt, parsep=0pt}

\theoremstyle{acmdefinition}
\newtheorem{defn}{Definition}

\begin{document}

\title[FreeForm Harness]{FreeForm Harness: A Mobile-Native Agent Harness for Open and Composable Extensions}

\author{Bo Zhang}
\authornote{Corresponding author.}
\affiliation{%
  \institution{Amap, Alibaba Group}
  \city{Beijing}
  \country{China}}
\email{cleroman@163.com}

\author{Pengbo Zhang}
\affiliation{%
  \institution{Amap, Alibaba Group}
  \city{Beijing}
  \country{China}}

\author{Jiaxin Jiao}
\affiliation{%
  \institution{Amap, Alibaba Group}
  \city{Beijing}
  \country{China}}

\author{Yue Liu}
\affiliation{%
  \institution{Amap, Alibaba Group}
  \city{Beijing}
  \country{China}}

\begin{abstract}
Mobile phones already contain the capabilities required by many everyday agent
tasks, but current mobile agents expose them only through monolithic pipelines
with no open recombination point. Users and third parties therefore cannot
publish, install, and compose new behaviors, and the personalized long tail of
daily-life requests remains largely unserved. We argue that the missing piece
is not capability but substrate. This is a Web infrastructure problem: the
phone is now the Web's dominant access point, yet the Web's defining property,
open publication and composition under decentralized authorship, stops at the
agent harness.

We present \textbf{FreeForm Harness (FFH)}, a mobile-native agent harness for
open and composable extensions. FFH separates what must always be present from
what should remain freely replaceable: a frozen base and a permanence-aware
required layer absorb mobile-native constraints such as location, proactive
triggering, and scarce channels, while a uniform plugin contract keeps
independently authored extensions composable, checkable, and replaceable over a
frozen socket set. FFH also extends openness from third-party publication to
end-user authorship through a tiered commissioned-authoring path: when no
existing extension matches a user's intent, a Web coding agent produces a
candidate artifact that is still validated and admitted on device under the
same contract discipline as any other extension.

An Android prototype realizes FFH across device, market, and cloud. Evaluation
establishes three results: the claimed architectural boundaries are enforced in
the running system; realistic daily-life requests fit the declared composition
space without requiring it to widen; and the full path from user intent to
on-device execution closes through retrieval or commissioned authoring while
final trust decisions remain device-local. FFH thus advances the agentic Web from
``everything is a plugin'' toward ``everyone is a developer'' on mobile.
\end{abstract}




\maketitle

\hypersetup{pdfauthor={Bo Zhang, Pengbo Zhang, Jiaxin Jiao, Yue Liu}}

\section{Introduction}
\label{sec:intro}

\begin{figure*}[t]
\centering
\includegraphics[width=\textwidth]{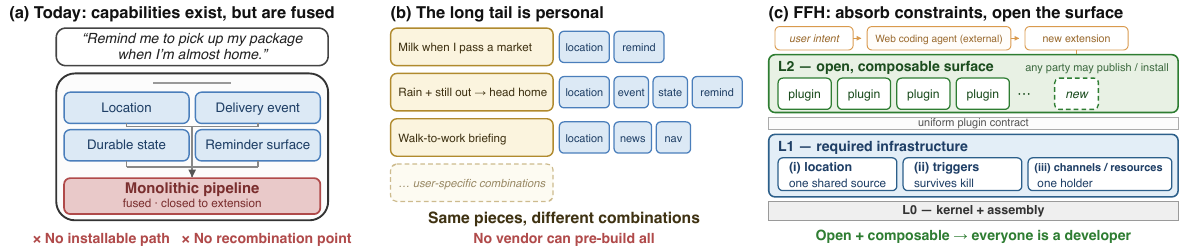}
\caption{Conceptual overview. (a)~The pieces needed for the motivating request
already coexist on the phone, but current agents fuse them into a monolithic
pipeline with no installable path or recombination point for users or third
parties. (b)~The personalized long tail recombines the same pieces in
user-specific ways, so no vendor can pre-build it. (c)~FFH absorbs location,
proactive triggering, and scarce channels or resources into required
infrastructure $L_1$, leaving an open, contract-governed $L_2$ where any party
may publish or install extensions, including user-commissioned ones via a Web
coding agent.}
\label{fig:concept}
\Description{Three-panel conceptual overview. Panel (a) shows that the
motivating request already depends on capabilities present on the phone, but
that current mobile agents expose them only through a fused pipeline with no
installable recombination point for the user or a third party. Panel (b) shows
several daily-life requests built from overlapping capability pieces, conveying
that the long tail comes from different user-specific combinations rather than
from entirely different primitives. Panel (c) shows FFH's layered response: a
frozen base, a required layer that absorbs location, triggering, and scarce
channel constraints, a uniform plugin contract, and an open extension layer on
which existing or newly authored plugins can be installed.}
\end{figure*}

\emph{``Remind me to pick up my package when I'm almost home.''}
A phone already has the needed pieces. It can track location, receive delivery
notifications, keep state for hours or days, and show a reminder through a
banner, widget, or badge without interrupting the foreground app. What is
missing is not capability but an \emph{open recombination point}. Today's
mobile agents expose these capabilities only through fused pipelines. Users
cannot install a new behavior that combines them differently, and third parties
cannot publish an extension that others can install.

This gap reflects a broader long tail of everyday requests: remind me to buy
milk when I pass a supermarket; if rain is forecast and I am still outside,
tell me to head home; fill my walk to work with financial news interleaved with
navigation prompts. These requests reuse many of the same building blocks, but
combine them differently for different users. No vendor can pre-build this long
tail, because it is precisely the space of user-specific combinations.

The problem is therefore architectural. More and more value now reaches the
user through Web services, but it is finally combined and acted on at the
device edge. The phone is the Web's dominant access point, yet the Web's
defining property, open publication and composition under decentralized
authorship, stops at the agent harness. Current mobile agent systems provide no
open point at which a third party can publish an extension, have it discovered,
verified, admitted, and executed on another user's device without vendor
permission. The missing piece is not simply a better mobile assistant. It is a
mobile harness that carries Web openness all the way to on-device agent
execution.

Two requirements follow. First, the extension surface must be \textbf{open}:
parties other than the vendor, in principle including the user, must be able to
author new behaviors. Second, composition must remain \textbf{general}: because
the long tail comes from recombining shared capability pieces in user-specific
ways, the architecture cannot hardwire only a small set of favored
combinations. Mobile agents therefore need \textbf{open and composable
extensions}.

Current systems still fall short of both. End-to-end mobile GUI agents such as
AppAgent \citep{zhang2023appagent}, Mobile-Agent-v3.5
\citep{xu2026mobileagentv35}, and MobiAgent \citep{zhang2025mobiagent} package
capability inside fused pipelines. Some are modular internally, but their parts
are not externally installable or replaceable at runtime. Siri and Google
Assistant do expose selected mobile behaviors, including location-based
reminders, but only as built-in features. More registry-oriented systems move
closer to modularity, but authorship remains developer-centered and admission
remains tied to a managed registry rather than an open extension surface. On
the desktop, DeepSeek Harness and Cordis \citep{shi2026composability} show an
important alternative: an agent can be organized as a plugin substrate rather
than as a monolithic assistant. But the mobile case is harder. The long tail
here is personal, context-sensitive, and shaped by user-specific choices of
trigger, data source, output modality, and threshold. On mobile, the plugin-substrate line should 
extend from \emph{everything is a plugin} toward \emph{everyone is a developer}: 
users should be able to state intent, have the
system commission the missing extension on their behalf, and still admit the
result under the same on-device contract discipline as any other plugin.

The difficulty is that mobile openness cannot come from making everything
optional. Three constraints become central once the authoring surface
opens: \textbf{location and mobility}, \textbf{proactive triggering}, and
\textbf{scarce channels and exclusive resources}. Location is a changing
execution context; trigger commitments must survive process death; and
independently authored extensions may compete for the same presentation slot or
platform-level exclusive resource. If each extension handles these constraints
itself, the result is duplication, interference, and lost continuity.
Mobile openness therefore requires always-present infrastructure that absorbs
these constraints once so that behaviors above it can remain open.

\begin{figure*}[t]
\centering
\includegraphics[width=\textwidth]{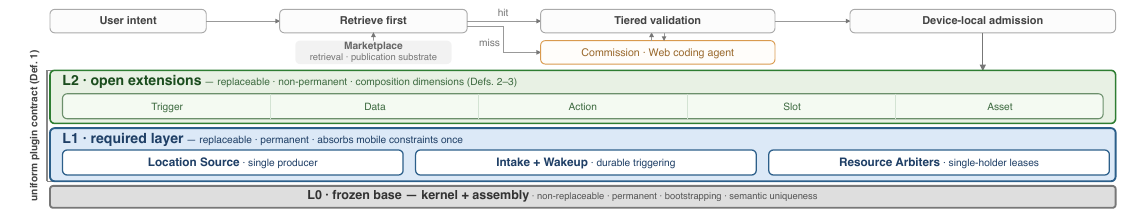}
\caption{Architecture overview. FFH comprises a frozen base $L_0$, a
replaceable-but-permanent required layer $L_1$, and an open extension layer
$L_2$. The uniform plugin contract (Definition~\ref{def:descriptor}) spans $L_1$ and $L_2$, fixing
the plugin-facing boundary with implementations replaceable. $L_1$ absorbs
location, proactive triggering, and scarce-resource contention via
always-present authorities; $L_2$ supports composition over trigger, data,
action, slot, and asset dimensions (Definitions~\ref{def:space}--\ref{def:orthogonal}). Top: the retrieval-first
authoring path---retrieve on hit, commission on miss, then tiered validation
and device-local admission under one discipline.}
\label{fig:architecture}
\Description{Architecture overview of FFH. The figure shows a layered harness
with a frozen base L0, a required layer L1, and an open extension layer
L2, all coupled through a uniform plugin contract. The required layer is
shown as the always-present infrastructure that absorbs location, proactive
triggering, and scarce-resource constraints. The open layer is shown as the
space in which plugins compose across five dimensions: trigger, data, action,
presentation slot, and asset. Above the layers, the figure shows the
retrieval-first authoring path: user intent first searches the marketplace,
retrieval misses commission a Web coding agent, returned artifacts pass
tier-appropriate validation, and admitted artifacts enter the open layer.}
\end{figure*}

We present \textbf{FreeForm Harness (FFH)}, a mobile-native agent harness for
\textbf{open and composable extensions}. FFH separates what must always be
present from what should remain freely replaceable through three layers: a
frozen base $L_0$, a replaceable-but-permanent required layer $L_1$, and an
open extension layer $L_2$ for independently authored plugins. A uniform plugin
contract spans $L_1$ and $L_2$, keeping extensions composable, checkable, and
replaceable, while $L_1$ absorbs location, proactive triggering, and
scarce-resource contention through always-present authorities.
FFH also extends openness from third-party publication to end-user
commissioning. Users can install, compose, and replace extensions directly.
When no existing extension matches a user's intent, the user can state that
intent in natural language, a Web coding agent can generate a candidate
artifact, and FFH validates and admits that artifact on device under the same
contract discipline as any other extension. Commissioned authoring is not
a privileged bypass, but another path into the same extension surface. Figure
\ref{fig:concept} gives the overview, and Figure
\ref{fig:architecture} previews the architecture. Our contributions are:

\begin{enumerate}
  \item \textbf{A permanence-aware architecture for open mobile agents.}
  We identify the missing architectural boundary for mobile openness and
  introduce a three-layer harness that separates a frozen base, a replaceable
  but permanent required layer, and an open extension layer.

  \item \textbf{Constraint absorption as the condition for mobile openness.}
  We show that location, proactive triggering, and scarce-resource contention
  must be absorbed into always-present authorities, allowing open extensions to
  remain composable above them under a uniform contract.

  \item \textbf{A unified extension lifecycle from publication to end-user
  commissioning.}
  We present a retrieval-first, tiered authoring model that subjects published
  and AI-generated extensions to the same on-device validation and admission
  process.

  \item \textbf{An end-to-end Android realization and evaluation.}
  We implement FFH across device, market, and cloud, and show that its
  architectural boundaries are enforced in the running system, its declared
  composition space covers realistic daily-life requests without widening, and
  its extension pipeline closes from user intent to admitted on-device
  execution while final trust decisions remain device-local.
\end{enumerate}

\section{The FreeForm Harness Architecture}
\label{sec:arch}

\subsection{Three-Layer Structure}
\label{sec:layers}

An open mobile harness must satisfy two requirements from
Section~\ref{sec:intro}: the extension surface must be open, and composition
must remain general. On a phone, however, some guarantees cannot be left to
optional extensions: location is a changing execution context, trigger
commitments must survive process death, and independently authored extensions
may compete for scarce channels or exclusive resources. The architectural
question is therefore not whether these constraints exist, but where they
belong.

Two simple designs fail. A monolithic agent keeps all critical machinery
present, but gives up openness. A fully dynamic plugin system keeps everything
replaceable, but cannot guarantee that the components needed for location,
triggering, or resource control will still be present in every execution. The
difficulty is not a fundamental trade-off between openness and control; it
comes from conflating two different properties.

A component is \textbf{replaceable} if it conforms to a declared interface,
states its dependencies explicitly, and can be swapped without changing its
neighbors. A component is \textbf{permanent} if the architecture guarantees
that some implementation of it is present in every execution after bootstrap.
Replaceability concerns whether an implementation may change; permanence
concerns whether the role may disappear.
This distinction yields FFH's three-layer structure. FFH separates the harness
into non-plugin machinery $\mathcal{K}$ and plugins $\mathcal{P}$. The frozen
base is $L_0 = \mathcal{K}$, which is non-replaceable and permanent. It
bootstraps the plugin system and fixes semantics that cannot vary at runtime.
Among plugins, the replaceable-yet-permanent subset is
$L_1 \subseteq \mathcal{P}$, the required layer. Its members may be replaced
by other conforming implementations, but some implementation of each required
role must always be present. The remaining plugins form
$L_2 = \mathcal{P} \setminus L_1$, the open extension layer. These plugins are
replaceable and non-permanent: they may be installed, removed, and recombined
freely.

\begin{table*}[t]
\centering
\caption{Layer partition criteria.}
\label{tab:partition}
\footnotesize
\setlength{\extrarowheight}{2pt}
\begin{tabularx}{\textwidth}{@{}>{\raggedright\arraybackslash}p{1.35cm}
                              >{\raggedright\arraybackslash}p{1.9cm}
                              >{\raggedright\arraybackslash}p{2.7cm}
                              >{\raggedright\arraybackslash}X@{}}
\toprule
Layer & Replaceability & Permanence & Membership criterion \\
\midrule
$L_0$ (kernel) & Non-replaceable & Permanent: absence either halts the plugin
system or violates a harness invariant & At least one of: \textbf{(1)
Bootstrapping}: without $c$, the plugin system cannot operate or plugins cannot
interpret each other's declarations. \textbf{(2) Semantic uniqueness}: the
harness admits only one legal semantics for $c$. For normative assets like
rules, what is fixed is the rule text or declared meaning, not whichever
executor applies it. Allowing several runtime-selectable semantics
here would turn a harness invariant into a mere default. \\
\addlinespace[2pt]
$L_1$ (required) & Replaceable & Permanent: absence breaks at least one
always-required wing & Both of: \textbf{(1) Three-wing absence test}: without
$c$, at least one of \textbf{safety} (a path exists on which the user may be
harmed without awareness, including privacy loss, financial cost, or data
loss), \textbf{authoring} (the intent--generate--validate--apply loop cannot
proceed), or \textbf{continuity} (the agent ceases to be continuously present:
state is not durable, process revival is impossible, or no unique source of
external-world facts remains) is violated. \textbf{(2) Non-bootstrapping}:
the plugin system itself remains operable without $c$. \\
\addlinespace[2pt]
$L_2$ (open) & Replaceable & Non-permanent: freely & Conforms to the
plugin contract, and its absence violates none of the three wings of the $L_1$
criterion. \\
\bottomrule
\end{tabularx}
\end{table*}

Table~\ref{tab:partition} operationalizes this partition by distinguishing
bootstrapping and semantically unique elements ($L_0$),
permanent-but-replaceable roles ($L_1$), and the removable remainder ($L_2$).
The key architectural move is $L_1$. Because no plugin on the open surface can
be assumed present, FFH assigns permanence only to infrastructure whose
absence breaks at least one always-required wing---\textbf{safety},
\textbf{authoring}, or \textbf{continuity}---as defined in
Table~\ref{tab:partition}.

The partition is complete by construction: $\mathcal{K}$ and $\mathcal{P}$ are
disjoint and exhaustive; $L_0$ contains the bootstrapping and semantically
unique elements; every plugin then lands in exactly one of $L_1$ or $L_2$
under Table~\ref{tab:partition}. Where several implementations realize the same
required role, permanence attaches to the role rather than to any one
implementation. Appendix~\ref{app:membership} records the complete membership
witness. This partition lets FFH remain open without giving up mobile
guarantees: $L_0$ and $L_1$ carry what must persist, while $L_2$ remains freely
installable, removable, and replaceable above them. Section~\ref{sec:contract}
defines the uniform contract that governs all plugins, and
Section~\ref{sec:absorption} shows how the required layer absorbs the mobile
constraints that openness cannot safely leave to chance.

\subsection{Uniform Plugin Contract}
\label{sec:contract}

The three-layer split answers what must remain present. The next question is
how components across those layers interoperate without knowing each other's
implementations. FFH therefore uses one uniform declaration surface for all
plugins. It inherits the plugin model of Cordis
\citep{shi2026composability}, where compatibility is checked over declarations
rather than discovered through runtime calls. FFH extends that model for mobile
by lifting into the declaration surface facts that would otherwise remain
implicit until runtime, including hook ownership and interception, event
emission authority, configuration type, and activation policy. The purpose of
Definition~\ref{def:descriptor} is not to formalize every runtime detail, but to make
interaction-relevant facts explicit enough for pre-execution compatibility
checks and coherent replacement.

\begin{defn}[Plugin Descriptor]
\label{def:descriptor}
A plugin descriptor is the plugin's complete declaration surface. Formally, a
descriptor is a 9-tuple
\[
d = \langle R, O, P, H_o, H_b, H_w, E, C, \alpha \rangle
\]
where:
\begin{itemize}
  \item $R \subseteq \text{ServiceKey}$ is the set of required services.
  \item $O \subseteq \text{ServiceKey}$ is the set of optional services.
  \item $P \subseteq \text{ServiceKey}$ is the set of provided services.
  \item $H_o \subseteq \text{HookPoint}$ is the set of owned hook points.
  \item $H_b \subseteq \text{HookPoint}$ is the set of observed hook points.
  \item $H_w \subseteq \text{HookPoint}$ is the set of wrapped hook points.
  \item $E \subseteq \text{EventPrefix}$ is the set of event prefixes the
  plugin may emit.
  \item $C$ is the type of runtime configuration accepted by the plugin.
  \item $\alpha \in \{\text{eager}, \text{lazy}\} \cup
  (\{\text{event}\} \times 2^{\text{EventPrefix}})$ is the activation policy.
\end{itemize}
Well-formedness requires $R \cap O = \emptyset$. Each optional dependency in
$O$ must have a declared default behavior. An instance may omit a key from the
class's $R$ only when its configuration selects execution paths that never
invoke it; the sandboxed trial enforces this premise.
\end{defn}

Because interaction-relevant facts are
declared in bounded form, three guarantees follow.
\textbf{Static checkability:} assembly compatibility can be
checked before code executes. No hook point may be owned by more than one
plugin; no hook point may be observed or wrapped without an owner; wraps may
short-circuit whereas observes may not; and every event-triggered activation
must name only prefixes that some plugin may emit. \textbf{Provider
uniqueness:} each service key admits at most one provider across all instances,
eliminating contract-surface ambiguity. \textbf{Capability narrowing:} for
declarative and scripted payloads, the runtime mediates interaction through the
descriptor; native Tier-3 payloads are bounded by provenance,
admission, test-base evidence, and crash containment.

Some facilities (e.g., clock, dispatchers, configuration, package-local
assets, and private storage) are ambient runtime provisions rather than
per-plugin grants. Uniform provisioning means a common interface, not shared
access: namespaces remain isolated and event emission remains bounded by
declared prefixes.

At the instance level, FFH allows only narrowing relative to the plugin class.
Let $d_c$ be a class descriptor and $d_i$ an instance descriptor. Then
\[
d_i \preceq d_c \iff R(d_i) \subseteq R(d_c) \;\wedge\; O(d_i) \subseteq O(d_c)
\;\wedge\; \alpha(d_i) \succeq \alpha(d_c).
\]
Here $\succeq$ is the admission ordering over trigger generality, with
$\text{eager} \prec \text{lazy} \prec \text{event}$. When both class and
instance are event-triggered, the instance's prefix set must be contained in
the class's declared prefix set. When the class is eager or lazy, any
event-triggered instance must draw its prefixes from $E(d_c)$.

Two runtime commitments keep a changing system coherent. \textbf{Spatial
coherence:} dependencies in $R$ and $O$ are resolved against the live registry,
and an absent required service suspends the dependent rather than failing it.
\textbf{Temporal coherence:} every contribution registers as a reversible
effect, so unloading a plugin leaves no residue in registries, event buses, or
schedulers.

\subsection{Constraint Absorption}
\label{sec:absorption}

The contract makes plugin structure explicit, but it does not govern
shared mobile resources. If no always-present authority controls such a
resource, each plugin manages it locally and none can predict or reconcile the
rest. On a phone, this problem appears most sharply in location, proactive
triggering, and scarce channels or exclusive resources. FFH addresses these by
placing each contested resource under an always-present authority in $L_1$.
Plugins declare only local needs and effects, but the global guarantees
those declarations rely on are enforced once in required infrastructure. 
These authorities are plugins rather than hardwired kernel
code: their roles are permanent, though their implementations remain
replaceable.

\textbf{Location $\rightarrow$ Location Source.}
If location is left to open extensions, each may start its own producer with
its own sampling policy, wakelocks, and accuracy requests. Battery cost then
scales with the number of consumers, and spatial semantics fragment across
policies. FFH instead enforces a \textbf{single-producer invariant}: one $L_1$
plugin is the sole producer of location facts for the harness. It governs
sampling and accuracy, tags facts with their coordinate reference system, and
normalizes them before exposure to consumers.

\textbf{Proactive triggering $\rightarrow$ Intake Dispatcher + Background
Wakeup.}
Trigger commitments must survive process death, so they cannot live only inside
transient plugin state. Left to open extensions, triggering loses both
\textbf{durability} and \textbf{integrity}: no common baseline records what
commitments should exist after revival, and orphaned registrations or duplicate
wake paths have no single authority to reconcile them. FFH absorbs this
constraint with two $L_1$ authorities. The \textbf{intake dispatcher} governs
arrivals while the process is alive: it normalizes external arrivals into log
facts, applies queue-or-preempt policy, and wakes the agent loop. The
\textbf{background wakeup} authority governs persistence across process death:
it owns the durable registration ledger for declared conditions and reconciles
it against the event log on revival, while the host exposes platform scheduling
handles through a reserved seam. The key invariant is \textbf{write-ahead
registration}: every registration or cancellation of a scheduling handle must
first be recorded in the log. The log is authoritative at revival.
Reconciliation is idempotent: handles without log backing are discarded, logged
registrations missing a live handle are re-established, and logged
cancellations backed by a live handle are revoked. FFH does not override
platform power policy; its claim is narrower: trigger commitments remain
durably represented and mechanically recoverable across process death, subject
to the wake paths the platform permits.

\textbf{Scarce channels and exclusive resources\hspace{0pt}
$\rightarrow$\hspace{0pt} Exclusive-\hspace{0pt}Resource Arbiter.}
Some resources must be held rather than merely read, including
platform-arbitrated resources such as audio focus and camera, and FFH
single-writer leases over logical resources such as a presentation slot or a
location subscription. Left to open extensions, such resources produce
interference that no plugin can resolve alone. FFH absorbs this constraint
through an $L_1$ \textbf{exclusive-resource arbiter}. Each governed resource is
represented as a named lease with at most one holder. Plugins claim resources
through the contract surface and declare how they should be treated when the
resource is busy. The arbiter resolves contention with priority, epoch-based
invalidation of stale holders, and a three-state disposition of
\emph{interrupt}, \emph{defer}, or \emph{discard}. Requests that need several
leases acquire them through one arbiter-coordinated transaction, so either all
required leases become valid together or none do.

Taken together, these authorities are the operational content of $L_1$: open
plugins declare local needs and effects, while the mobile guarantees are
enforced once in always-present infrastructure.

\subsection{Composition Model}
\label{sec:composition}

With mobile constraints absorbed into $L_1$, FFH composes $L_2$ behaviors by
configuration, following Cordis and DeepSeek Harness
\citep{shi2026composability}: an interpreter class fixes the execution
envelope, and a typed runtime specification fixes behavior within it.
For the reactive mobile behaviors studied here, a specification must capture
five choices: \emph{when} it fires, \emph{what} information it draws on,
\emph{what} effect it produces, \emph{where} that effect appears, and
\emph{what} static content it carries. FFH therefore projects each
specification into five dimensions: trigger, data, action, presentation slot,
and asset. Architecturally, these five dimensions are the declarative envelope
for the reactive-mobile class studied here; \S\ref{sec:eval:composition}
confirms that every benchmark case projects into this envelope without
widening it.

\begin{defn}[Composition Space]
\label{def:space}
A plugin specification $s$ is projected into a five-dimensional composition
space
\[
\pi(s) \in \mathcal{B} = \mathcal{T} \times \mathcal{D} \times \mathcal{A}
\times \mathcal{S} \times \mathcal{As},
\]
whose dimensions correspond to trigger, data, action, presentation slot, and
asset.
\end{defn}

The point of this projection is architectural: many new behaviors differ only
in how these five choices are combined, not in requiring a new pipeline or
interpreter.

\begin{defn}[Dimension Orthogonality]
\label{def:orthogonal}
Fix a declarative cross-constraint set $\Xi$ over dimension values, represented
as a finite set of predicates stored in the contract's rule assets. For a fixed
interpreter class, two dimensions $X_i$ and $X_j$ of $\mathcal{B}$ are
orthogonal relative to $\Xi$ if changing an admissible value in one dimension,
while holding the others fixed, yields another admissible specification
whenever the resulting tuple does not violate $\Xi$, and if adding a new value
to either dimension requires no change to the interpreter class or to existing
specifications.
\end{defn}

In plain terms, one dimension can vary without rewriting interpreters or
existing specifications unless a declared cross-constraint forbids the
combination. This separates the \emph{candidate space} of combinations from the
\emph{admissible subset} selected by the interpreter envelope and $\Xi$. For a
finite snapshot, adding one new value to dimension $X_i$ expands the candidate
space by
\[
|\mathcal{B}'| - |\mathcal{B}| = \prod_{j \neq i} |X_j|.
\]
Architecturally, widening one dimension multiplies combinations through the
others without requiring new pipeline code.

The five dimensions do not exhaust $L_2$: native algorithms remain code-based
plugins, and new engines remain outside the declarative envelope. Their value
is analyzability: any admitted behavior must still project its triggers,
dependencies, actions, and resource claims into the contract surface of
Definition~\ref{def:descriptor} and the composition dimensions of
Definition~\ref{def:space}. Declarative growth therefore comes from widening
orthogonal dimensions; behaviors outside that envelope arrive as authored
artifacts under the same contract.

\subsection{AI-Mediated Authoring}
\label{sec:authoring}

A marketplace can accumulate extensions from many authors, but it cannot cover
the full long tail of Section~\ref{sec:intro}. FFH therefore extends openness
from third-party publication to end-user commissioning: when retrieval misses,
the system may create on the user's behalf, but the result still enters through
the same admission path as any other extension.

FFH selects among three tiers using two predicates: whether the
needed engine is already available ($G$), and whether the requested behavior is
expressible as a zero-logic specification over existing contracts and APIs
($Z$). If $G \wedge Z$, the request is Tier~1; if $G \wedge \neg Z$, Tier~2;
if $\neg G$, Tier~3. The lowest sufficient tier always wins. Before tier
selection, FFH returns \emph{unsupported} if the device lacks the needed
capability and \emph{unauthorable} if the request requires a platform
declaration the harness does not possess.

The tiers correspond to three artifact classes. \textbf{Tier~1} is a zero-logic
specification, i.e., a point in the five-dimensional composition space of
Section~\ref{sec:composition}. \textbf{Tier~2} is restricted script executed
inside the harness sandbox over engines already present on the device.
\textbf{Tier~3} introduces native capability not already available inside the
harness, as a package that implements an existing contract, adds a tool under
an admitted generic contract, or supplies a new engine.

The on-device agent is the requester, validator, and final acceptor; the Web
coding agent is an off-device commissioned developer; and the marketplace is
the retrieval and publication substrate. Validation occurs where the artifact's
runtime can exist and fail reversibly: Tier~1 and Tier~2 are validated entirely
on device, whereas Tier~3 additionally passes through a cloud test base of
disposable runtimes. Final admission remains device-local.

Validation certifies \emph{admissibility}, not semantic perfection: machine
checks establish contract conformance and trial safety, while semantic fit is
judged by the user on the running artifact. For Tier~1 and Tier~2, the order is
render $\rightarrow$ accept $\rightarrow$ admit. Tier~3 instead uses bounded
post-install trial on Android: after a green cloud report, the device installs
the package transactionally, records a trial deadline, and retains it only if
the user accepts; refusal or expiry rolls it back. For replacement, where
rollback would discard the incumbent, confirmation is requested before the
transaction.

Operationally, FFH is retrieval-first: it first searches the marketplace and
commissions the Web coding agent only on a miss, using a brief derived from the
intent, schema, rules, and templates. Candidate artifacts iterate through
tier-appropriate validation until one is admitted or the request is rejected.
Publication back to the marketplace is optional and grants no exemption: later
downloads traverse the same admission pipeline, minus generation. This keeps
commissioned authoring bounded by existing contracts and preserves the
no-bypass admission rule.

\section{Android Realization}
\label{sec:impl}

FFH separates \emph{frozen mechanism} from \emph{mutable membership}, and the
Android realization preserves this split as a deployment boundary. The base
that makes plugins loadable, checkable, and runnable ships in the APK, while
plugin implementations arrive as signed packages and can change at runtime.
Thus, any capability expressible by the frozen contracts evolves through
replaceable packages rather than APK updates. This is the concrete form of FFH's central
claim: the harness stays fixed where trust and continuity concentrate, while
capability evolves through replaceable membership.

\begin{figure}[t]
\centering
\includegraphics[width=\columnwidth]{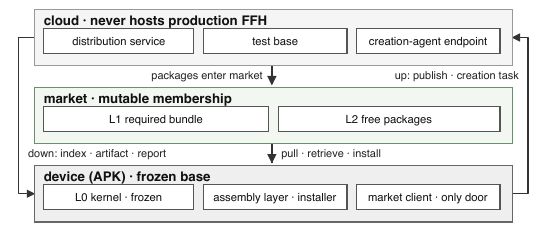}
\caption{Deployment boundary. The frozen base ships in the APK, mutable
membership flows through the market, and the cloud provides only supporting
services for distribution, Tier-3 verification, and commissioned creation.
Production execution and final admission remain device-local.}
\label{fig:deployment}
\Description{Three stacked bands for the cloud, the market, and the device,
each with aligned chips: distribution service, test base, and creation agent;
the L1 required bundle and the L2 free packages; and the L0 kernel, the
assembly layer, and the market client as the bootstrap channel to the market.
Two centered arrows on one axis cross the band gaps: reviewed packages are
listed into the market, and the device pulls, retrieves, and installs through
its market client. A left rail carries the signature index, artifacts, and
reports down to the device, fail-closed; a right rail carries publish and
creation tasks up to the cloud; both rails bypass the market.}
\end{figure}

On Android, this split yields three administrative domains
(Figure~\ref{fig:deployment}). The \textbf{device} is the only production
runtime: it hosts the $L_0$ kernel, the assembly layer, and the built-in market
client, and it is the only place where final admission is decided. The
\textbf{market} carries mutable membership, including the $L_1$ bootstrap set
and all $L_2$ extensions. The \textbf{cloud} provides only supporting services:
package distribution, the Tier-3 test base, and the commissioned-authoring
endpoint. It never hosts a production harness. This split makes execution device-local and evolution package-based, and the
rest of this section shows how Android realizes that boundary.

\subsection{Frozen Base and Mutable Membership}
\label{sec:impl-base-market}

The APK ships mechanism, not evolving end-user capability. It contains the
$L_0$ kernel, the assembly layer that installs and binds plugins, and the
built-in market client that bootstraps mutable membership. The kernel provides
the registry, lifecycle machinery, event-log semantics, contract library, and
validation rules. The assembly layer handles manifest adjudication, loading,
signature checks, contract registration, and the host UI shell. The market
client must ship in the APK because the device needs a bootstrap channel to
retrieve packages that are not yet present locally.

Two invariants govern the APK. First, the \emph{empty-socket invariant}: the
APK ships contracts and mechanism, but no evolving domain-capability
implementations. Second, the \emph{frozen-surface invariant}: contract
signatures, manifest and log formats, and rule assets change only with an APK
release. Together, these invariants make the APK a stable execution substrate
rather than a carrier of changing end-user capability.

Mutable membership flows through the market. The $L_1$ bootstrap set supplies
the required roles that complete the harness after installation, while $L_2$
contains optional extensions. Operationally, $L_2$ artifacts may be added or
removed freely, whereas $L_1$ artifacts are admitted only as same-function
replacements under stricter verification, including re-checks of the replaced
role's constraints.

All market packages share one envelope: identity, signature, dependency
declarations, and a typed payload. Payloads may be declarative specifications,
sandboxed scripts, or native dex packages. Trust is layered rather than
centralized: creator signatures are checked before load, the device consults a
signature index with revocation state, and users may additionally pin creator
identities directly. Admission remains device-local and fails closed when the
required trust evidence is unavailable. Once a package has been admitted,
execution remains offline-capable from the admitted snapshot; trust checks gate
membership change rather than steady-state execution. For native artifacts, installation 
is identity-bound, re-verified at load time, and retained only after durable acceptance.

\subsection{Cloud Support Services}
\label{sec:impl-cloud}

The cloud is the only administrative domain that never hosts a production
harness. Its role is deliberately narrow: it supports distribution and
creation, but does not participate in steady-state execution or final
admission.
It provides three services. The \emph{distribution service} stores market
packages and publishes trust metadata such as the signature index and
revocation state. The \emph{test base} is used only for Tier-3 artifacts,
running candidate native packages in disposable Android runtimes before they
may be delivered to a device. The \emph{creation-agent endpoint} serves
commissioned authoring requests as bounded tasks rather than as an interactive
remote runtime. Because the cloud supports only distribution and creation, the device remains
offline-capable for local lifecycle operations once packages have been
retrieved and admitted. Connectivity is required only for inherently remote
tasks such as fetching packages, consulting remote trust state, or
commissioning new artifacts.

\subsection{Device-Local Runtime and Admission}
\label{sec:impl-run}

To host a changing plugin set above a frozen base, FFH keeps recovery,
membership change, and admission coherent under process death and replacement
through three device-local mechanisms.

\textbf{Replay-based recovery.} On cold start, the device reconstructs runtime
state by replaying the event log against the installed membership; the log
remains authoritative across process death, including the write-ahead trigger
registrations of Section~\ref{sec:absorption}.

\textbf{Transactional membership change.} Installation, replacement, and
removal are realized as manifest transactions. A transaction validates the new
membership, prepares the change, and rolls back on failure before the old state
is irreversibly released. Dependents therefore suspend rather than error, and a
partially withdrawn membership state is never exposed.

\textbf{Uniform admission pipeline.} Every artifact that introduces a new
behavior passes through the same admission ladder: static checks, a
tier-appropriate trial where the runtime permits one, and user acceptance on
the running artifact. Replacing an $L_1$ implementation additionally re-runs
the replaced role's constraint suite before commit. Tier~1 and Tier~2 artifacts
are validated entirely on device; Tier~3 additionally requires a green report
from the cloud test base before bounded on-device trial and final acceptance.

These mechanisms are the implementation counterpart of
Section~\ref{sec:arch}. Continuity, membership change, and final trust remain
anchored on device even though publication, retrieval, and commissioned
authoring are open.

\section{Evaluation}
\label{sec:eval}

We evaluate FFH as an architecture for open and composable mobile extensions on
the Android prototype of \S\ref{sec:impl}. All experiments run on a Xiaomi 14
Pro under Android 16 with the live cloud services of \S\ref{sec:impl}. Our
goal is not to benchmark a particular coding model, but to test whether FFH
(i) enforces its claimed guarantees under open extension, (ii) covers a broader and more
stable reactive-mobile composition class than simpler automation abstractions,
(iii) closes the loop from user intent to admitted on-device behavior, and
(iv) does so at bounded mobile cost.

\textbf{Workload and baselines.}
We construct a 50-case benchmark of reactive-mobile requests from anonymized
daily-life requests and public trigger-action recipes
\citep{ur2016trigger,ur2014tap}. Cases are included only
when they admit a concrete delivery oracle; explicit unsupported,
unauthorable, and principled-refusal cases are also included so that honest
non-closure is measured rather than hidden. Principled non-admission is
therefore counted as honest non-closure, not as silent failure, since FFH
claims bounded openness under device-local admission rather than unconditional
closure. We compare \emph{FFH-full} against three baseline families: \emph{internal
weakened variants}, which remove one guarantee-enforcing mechanism while
holding the rest of the harness fixed; \emph{pipeline baselines}, which remove
one openness stage such as commissioned authoring, preview, or device-local
final acceptance; and an \emph{automation baseline}, a concrete recipe-style
trigger-condition-action substrate over the same device and delivery oracles.
The automation baseline provides a fixed workflow schema with recipe-scoped
and shared state, plus bounded inline scripting, but lacks reusable providers,
shared durable log queries, first-class slot or asset dimensions, same-function
replacement, commissioned authoring, and device-local admission.

\subsection{Architectural Guarantees and Necessity}
\label{sec:eval:guarantees}

We test whether FFH enforces the guarantees motivating its required
layer and uniform contract: safe assembly, mediated boundaries, coherent
dependency handling and replacement, transactional membership change,
single-source location, durable triggering, single-holder arbitration, and
the frozen-surface split (Table~\ref{tab:eval-guarantees}).

FFH-full enforced all tested contract boundaries. Malformed artifacts were
rejected before activation, conforming controls were admitted, mediated
over-reach was refused, required-provider removals suspended
dependents crash-free, optional-provider removals fell back to declared
defaults, provider replacements rebound through the live registry, and unload
left no residue in harness-owned state. Interrupted installs rolled back
cleanly, and audited turns matched the independently specified log projection.
These results show FFH's contract is operational rather than merely
descriptive. The required authorities likewise preserved their invariants.
Location remained single-source and lease-governed under concurrent consumers;
write-ahead reconciliation preserved trigger commitments across kill--restart;
and the arbiter preserved single-holder behavior under contention and injected
OS seizure. A released-package audit found no frozen-surface split
violation: the APK contained no undeclared domain capability, $L_1$ uninstalls were
refused, bootstrap recovery succeeded after data wipe, and same-function
substitutions passed their re-run constraint suites.

The weakened variants show that these properties do not survive removal of the
corresponding mechanisms. Disabling host-tracked rebinding and reversible cleanup produced stale bindings
and residual harness state. Removing write-ahead reconciliation produced
orphaned handles and missing re-established registrations after kill--restart.
Replacing required authorities with plugin-local best-effort coordination lost
single-source location semantics and single-holder resource behavior. Taken
together, these results support the architectural claim that FFH's guarantees
come from its contract and required authorities, not merely from having
plugins.

\subsection{Composition Coverage and Comparative Expressiveness}
\label{sec:eval:composition}

We next ask whether FFH's declared composition envelope is broad enough to
matter and broader than a simpler recipe abstraction. We evaluate three
properties: \emph{adequacy}, \emph{stability}, and \emph{comparative
expressiveness} against the automation baseline.
All 50 benchmark cases projected into FFH's five-dimensional composition space
without requiring a new dimension. Coverage was not confined to a narrow corner
of the model: 23 cases involved location or mobility, all 50 involved
proactive triggering, and all 50 touched at least one scarce presentation
channel or other exclusive resource. The benchmark also includes multi-step
coordinated actions, durable state suppression, guarded multi-source joins,
retrieval-ambiguous and preview-sensitive requests, and principled non-closure
cases. The interpreter audit found no case that required widening the declared
composition space by editing an existing interpreter.

\begin{table}[t]
\centering
\caption{Benchmark-level comparison.}
\label{tab:eval-coverage}
\footnotesize
\setlength{\extrarowheight}{1pt}
\begin{tabularx}{\columnwidth}{@{}>{\raggedright\arraybackslash}Xcc@{}}
\toprule
Metric & FFH & Automation \\
\midrule
Representable & 50 & 38 \\
Direct / workaround / not repr. & 50 / 0 / 0 & 15 / 23 / 12 \\
Closure / honest non-closure & 44 / 6 & 17 / --- \\
\bottomrule
\end{tabularx}
\end{table}

Table~\ref{tab:eval-coverage} summarizes the benchmark-level comparison, and
Table~\ref{tab:eval-main} in Appendix~\ref{app:eval} records the fuller probe
ledger behind these aggregates. FFH represented all 50 benchmark cases within
its declared composition envelope and closed 44/50 end to end, with the
remaining six terminating honestly as unsupported, unauthorable, or
principled-refusal outcomes. The automation baseline represented 38/50 cases in
total---15 directly, 23 only through workarounds, and 12 not at all without
adding a new capability class---and closed only 17/50 under the same delivery
oracles. The difference is therefore operational as well as representational:
workaround-based realizations were often too fragile or non-portable to satisfy
the benchmark's admitted execution criteria. The main deficits of the
automation baseline were log-derived absence checks, guarded multi-source
joins, slot-aware delivery, asset-bound behavior, and multi-step or stateful
fallback behavior.
We also tested substitution stability as same-function replacement over ten L1
slots with three implementations each, plus six tuples violating the declared
surface. Across the 36 substitutions, FFH realized 30 admissible variants
without interpreter modification and correctly refused six forbidden tuples
under the declared constraints or interpreter envelope. The automation baseline
was local in only 3/36 analogous substitutions; the remainder required workflow
rewrites, duplicated recipes, or were unsupported. Within the studied reactive-mobile
class, FFH therefore behaves as a stable composition envelope rather than a
collection of hardwired cases.

\subsection{End-to-End Openness}
\label{sec:eval:openness}

We evaluate FFH's openness claim as an end-to-end workflow from user intent to
admitted on-device behavior. For each benchmark case, FFH first attempts
retrieval from the marketplace; on a miss, it selects an authoring tier,
commissions the Web coding agent if needed, validates the returned artifact,
previews it in the tier-appropriate order, and admits it on device. We compare
FFH-full against three pipeline baselines: retrieval-only,
author-without-preview, and author-without-device-local-final-admission.

At the benchmark level, FFH closed 44/50 cases end to end, as summarized in
Table~\ref{tab:eval-coverage}. Retrieval closed 17 cases directly; 27 cases
entered commissioning; and six cases terminated honestly before artifact production,
spanning unsupported, unauthorable, and principled-refusal outcomes. Within the
commissioned set, four cases were routed to Tier~1, 22 to Tier~2, and one to
Tier~3. All 27 commissioned cases reached admission and executed successfully
in the main benchmark run. Revision burden remained bounded (mean 1.4, median
1, max 3), and preview rejected four machine-valid artifacts that were then
revised once and admitted.
These results show that FFH's openness is not reducible to reuse alone.
Retrieval-only closed only the 17 reuse hits, while retrieval-ambiguous cases
still required preview-mediated disambiguation or commissioned refinement.
Without preview, four machine-valid but semantically rejected artifacts would
have been admitted; without device-local final acceptance, they would have
become active without a final trust decision on device.
Against the automation baseline, which closed only 17/50 cases, FFH's pipeline
therefore expands closure beyond recipe-only automation.

To test robustness beyond a single clean pass, we re-ran a 10-case commissioned
subset five times each under the same coding-agent configuration. Across 50
runs, 46 reached admission and successful execution (92\%), with mean revision
count 1.6, stable tier routing in 48/50, and only limited intervention: 7/50
runs required one standardized restatement of intent, with no artifact editing,
prompt patching, or validation override. These results do not establish
robustness across arbitrary models or prompts, but they show that FFH's
openness pipeline is not a one-shot artifact of a lucky run and provide
strongest evidence for Tier~1 and Tier~2 closure, with Tier~3 demonstrated only
at feasibility level.

\subsection{Cost, Tradeoffs, and Limitations}
\label{sec:eval:tradeoff}

FFH's guarantees are not free, so we measure whether they come at
bounded rather than mobile-prohibitive cost. On the studied prototype, cold
start completed in 380\,ms median, replay-based recovery bounded above by
505\,ms, install in 178\,ms, replacement in 335\,ms, and unload in 19\,ms.
Trigger-to-action latency remained bounded at 1.7\,ms median for periodic and
event-triggered delivery and 0.2\,ms for geofence-triggered reminders, within
the warm process (wake-path costs excluded). Over an eight-hour
mixed-trigger run, FFH-full consumed 4.1\% battery, against 3.5\% for the
fully weakened variant and 3.2\% for the automation baseline. Thus FFH's stronger
guarantees come at measurable but bounded cost.

The evaluation has several limits. The benchmark is broad but not exhaustive; Tier~3
engine requests are underrepresented; the openness results depend in part on
the commissioned coding agent; all measurements were taken on one device model;
and we report neither a longitudinal non-programmer study nor an adversarial
publisher study. Our fail-closed claim is correspondingly narrow: when the loop
does not close before admission, FFH rejects, revises, or returns a structured
non-admission outcome rather than admitting unchecked behavior.

\section{Related Work}
\label{sec:related}

FFH relates to four nearby lines of work: mobile agents, plugin-based agent
substrates, end-user automation and skill reuse, and mobile automation
ecosystems with third-party extensions. Mobile GUI agents such as
AppAgent \citep{zhang2023appagent}, CogAgent \citep{hong2023cogagent}, the
Mobile-Agent line \citep{wang2024mobileagentv2,xu2026mobileagentv35}, and
MobiAgent \citep{zhang2025mobiagent} demonstrate increasingly capable
mobile-native interaction, but package capability inside end-to-end pipelines
rather than as externally installable runtime components. More system-oriented
platforms such as AIOS \citep{mei2024aios}, UFO3 \citep{zhang2025ufo3}, Aura
\citep{zou2026aura}, and NLAH \citep{pan2026nlah} move closer to modularity or
registry-based management, but do not establish FFH's combination of frozen
contracts, mutable plugin membership, device-local admission, and end-user
facing openness for mobile extensions. DeepSeek Harness and Cordis
\citep{shi2026composability} are FFH's closest architectural precursors: they
show that agents can be organized as plugin substrates with externalized
manifests and contract-mediated interoperability. FFH extends this line to mobile execution through a permanence-aware split
between frozen base and mutable membership, required-layer absorption of
mobile-native constraints, and device-local admission.

FFH also differs from prior work on reusable skills, trigger-action automation,
composition, location grounding, and mobile automation ecosystems with plugin
markets
\citep{wang2023voyager,wu2024oscopilot,li2026agentskillos,ur2014tap,ur2016trigger,li2019pumice,sutton1999options,ellis2021dreamcoder,kim2024llmcompiler,manvi2024geollm,chen2024mapgpt,yang2025contextagent,ning2025dima}.
These lines show that behaviors can accumulate, be recombined, and sometimes be
authored or installed by non-programmers, but not as a mobile-native harness
that simultaneously provides a uniform declarative contract, open runtime
installation and replacement, a permanence-aware required layer for location,
triggering, and scarce-resource control, device-local admission, and a unified
path from third-party publication to end-user commissioning. FFH is therefore
not simply another mobile assistant or automation tool, but a Web
infrastructure substrate for carrying open publication and composition to
on-device agent execution. Appendix~\ref{app:comparison} summarizes the
capability-level comparison.

\section{Conclusion}
\label{sec:conclusion}

This paper argues that openness on mobile is fundamentally an architectural
problem. FFH addresses it by separating what must persist from what should
remain open: a frozen base and permanence-aware required layer provide the
guarantees mobile execution cannot leave to chance, while a uniform contract
keeps extensions composable and replaceable above them. Our Android realization
and evaluation show that these boundaries are enforceable in the running
system, that the declared composition model covers realistic daily-life
requests without widening, and that the extension pipeline closes from user
intent to admitted on-device behavior while final trust remains device-local.
More broadly, FFH extends the plugin-substrate line of DeepSeek Harness and
Cordis \citep{shi2026composability} to the mobile agentic-Web setting. We hope
it can serve as a reference point for agent harnesses that remain frozen where
trust concentrates and fluid where creativity belongs.

\bibliographystyle{ACM-Reference-Format}
\bibliography{references}


\appendix

\section{Complete Layer Membership}
\label{app:membership}

Table~\ref{tab:membership} is the concrete witness for the partition claim of
Section~\ref{sec:layers}. It lists every catalogued $L_0$ member and every
required $L_1$ role together with the criterion or wing that places it there
and the consequence of its absence.
Assembly mechanisms that are non-plugin by construction, including market
intake validation, belong to $\mathcal{K}$ and are therefore not repeated as plugin rows here.
The built-in market client is the one $L_1$ role that ships inside the APK
rather than through the bootstrap set.

Membership is recorded per assembly row, but permanence in $L_1$ attaches to
the required \emph{role} rather than to any one implementation. Same-function
replacement must therefore preserve the service keys and guarantees that
justify the role's membership in $L_1$. The open remainder $L_2$ is not
enumerated.

\begin{table*}[t]
\centering
\caption{Complete layer membership witness for the partition of
Section~\ref{sec:layers}.}
\label{tab:membership}
\footnotesize
\setlength{\extrarowheight}{0pt}
\begin{tabularx}{\textwidth}{@{}>{\raggedright\arraybackslash}p{3.55cm}
                              >{\raggedright\arraybackslash}p{2.2cm}
                              >{\raggedright\arraybackslash}p{1.45cm}
                              >{\raggedright\arraybackslash}X@{}}
\toprule
Row ID & Member & Criterion / wing & Consequence of absence \\
\midrule
\texttt{kernel.plugin} & Plugin system kernel & Bootstrapping & No plugin can be registered, scheduled, or resolved \\
\texttt{kernel.declare} & Contract declaration primitives & Bootstrapping & Plugins cannot interpret each other's declarations \\
\texttt{kernel.runtime} & Runtime environment contract & Bootstrapping + semantic uniqueness & No uniform mediated runtime surface for execution \\
\texttt{kernel.faults} & Lifecycle fault isolation & Bootstrapping & One faulty install or stop can crash the whole process \\
\texttt{kernel.log} & Event-log semantics & Semantic uniqueness & No durable, replayable truth for recovery or audit \\
\texttt{kernel.invariants} & Invariant registry & Bootstrapping + semantic uniqueness & Invariants are neither registered nor checked \\
\texttt{kernel.contracts} & Contract library & Semantic uniqueness & No frozen socket vocabulary for cross-plugin collaboration \\
\texttt{kernel.rules} & Rule-asset catalog & Semantic uniqueness & Validation rules are absent at admission \\
\midrule
\texttt{agent.loop} & Agent loop & authoring & No turn executes; serving and authoring both stop \\
\texttt{effect.arbiter} & Effect arbitration & safety & Write effects execute without user adjudication \\
\texttt{exec.wrapper} & Execution wrapper & continuity & Calls run unbounded and uncancellable \\
\texttt{log.writer} & Event-log writer & continuity & State is not durably written across process death \\
\texttt{credential.vault} & Credential vault & safety & Keys and tokens lack a single protected entry point \\
\texttt{llm.route} & LLM routing & safety & Model egress and cost lack a single adjudication point \\
\texttt{permission.portal} & Permission portal & safety & Grants occur without a user-facing portal \\
\texttt{foreground.lifecycle} & Foreground lifecycle & continuity & Foreground presence and managed liveness are lost \\
\texttt{arbiter.voice} & Voice-channel lease & safety & Exclusive audio is held without arbitration \\
\texttt{arbiter.location} & Location subscription lease & safety & Exclusive location subscription is held without arbitration \\
\texttt{arbiter.camera} & Camera lease & safety & Exclusive camera access is held without arbitration \\
\texttt{arbiter.notification} & Notification-slot lease & safety & The presentation slot has no single holder \\
\texttt{arbiter.transaction} & Multi-lease transaction broker & safety & Multi-resource claims cannot be acquired all-or-nothing \\
\texttt{memory.store} & Memory store & continuity & Persistent memory tools and cross-turn facts are lost \\
\texttt{fault.notify} & Fault notification & safety & Failures occur without user awareness \\
\texttt{tool.registry} & Tool registry & authoring & Capabilities cannot be declared as tools to the loop \\
\texttt{session.service} & Session service & authoring & No session or derived views exist for interaction \\
\texttt{ui.registry} & UI registry & authoring & No slots exist for credential entry or permission grants \\
\texttt{recovery.coord} & Recovery coordinator & authoring & Incomplete turns are not recovered before progress \\
\texttt{creation.task} & Task state machine and artifact review & safety + authoring & Commissioned authoring lacks lifecycle and artifact adjudication \\
\texttt{retention.erase} & Retention erasure & safety & Deleted data may revive against policy \\
\texttt{platform.state} & Platform state portal & continuity & Platform facts lack a unique source \\
\texttt{intake.dispatcher} & Request intake & authoring & Arrivals drop without acceptance or logging \\
\texttt{location.source} & Location source & continuity & Location facts lack a unique producer \\
\texttt{wakeup.background} & Background wakeup & continuity & Declared conditions cannot revive the process \\
\texttt{egress.gate} & Egress adjudication & safety & Data egress lacks adjudication \\
\texttt{consent.ledger} & Consent ledger & safety & Consent facts are not durably recorded \\
\texttt{requirement.assemble} & Requirement assembler & authoring & Commissioned authoring cannot form a self-contained brief \\
\texttt{human.channel} & Human-interaction channel & authoring & Intents and confirmations lack a path \\
\texttt{model.provider} & Model provider & authoring & No completion path exists for intent understanding \\
\texttt{market.channel} & Built-in market client & authoring & Packages cannot be retrieved or published \\
\bottomrule
\end{tabularx}
\end{table*}

No $L_1$ member satisfies an $L_0$ criterion: removing any one degrades the
agent's guarantees, but does not prevent the plugin system itself from booting
or interpreting declarations. Conversely, the eight $L_0$ rows are exactly the
bootstrapping and semantically unique elements that define the harness. The
partition is therefore witnessed by eight $L_0$ members, the required $L_1$
roles above, and the open remainder $L_2$.

\section{Evaluation Detail Summary}
\label{app:eval}

This appendix records the minimum detail needed to interpret
Section~\ref{sec:eval}. In the main text, Table~\ref{tab:eval-coverage}
summarizes the benchmark-level comparison between FFH and the automation
baseline. Here we separate the appendix evidence by evaluation question rather
than collapsing all results into one ledger. Table~\ref{tab:eval-guarantees}
records the probe families and ablation outcomes behind the architectural
guarantees and necessity claims of \S\ref{sec:eval:guarantees}. Table
\ref{tab:eval-main} then summarizes the benchmark and closure outcomes used for
composition and openness, together with the robustness statistics reported in
the main text. 

More specifically, Table~\ref{tab:eval-guarantees} corresponds to
\S\ref{sec:eval:guarantees}; the \emph{Composition} rows of
Table~\ref{tab:eval-main} complement the benchmark-level comparison reported in
Table~\ref{tab:eval-coverage} of \S\ref{sec:eval:composition}; and the
\emph{Openness} rows of Table~\ref{tab:eval-main} correspond to
\S\ref{sec:eval:openness}, including the retrieval-first openness path and the
pipeline baselines used to isolate the contribution of commissioning, preview,
and device-local final acceptance. The appendix tables are intended as compact
audit trails for the headline numbers in the main paper rather than as
replacements for the narrative interpretation there.

\begin{table*}[t]
\centering
\caption{Architectural guarantees and necessity: enforcement probes and weakened-variant outcomes. Trial counts are not comparable across the two blocks: \emph{Guarantees} rows count probe scenarios, \emph{Ablation} rows randomized trials.}
\label{tab:eval-guarantees}
\footnotesize
\setlength{\extrarowheight}{1pt}
\begin{tabularx}{\textwidth}{@{}>{\raggedright\arraybackslash}p{2.2cm}
                              >{\raggedright\arraybackslash}p{3.2cm}
                              >{\raggedright\arraybackslash}p{2.4cm}
                              >{\raggedright\arraybackslash}p{2.0cm}
                              >{\raggedright\arraybackslash}X@{}}
\toprule
Part & Property / question & Probe family & Count / setup & Outcome \\
\midrule
Guarantees & Safe assembly and contract admission & Malformed artifacts / conforming controls & 14 / 4 & All malformed artifacts were rejected before activation; all conforming controls were admitted \\
Guarantees & Mediated contract boundary & Over-reach attempts & 6 & All mediated over-reach attempts were refused at the declared boundary \\
Guarantees & Coherent dependency handling & Required / optional removals & 10 / 10 & Required-provider absence suspended dependents without crash; optional-provider absence took declared defaults \\
Guarantees & Coherent replacement and unload & Provider replacements / unload trials & 5 / 20 & Live-registry rebinding was preserved; unload left an empty diff over harness-owned state \\
Guarantees & Transactional membership change and auditability & Interrupted installs / audited turns & 10 / 20 & All interrupted installs rolled back cleanly; model-visible context matched the independently specified log projection \\
Guarantees & Single-source location, durable triggering, and single-holder arbitration & Location / trigger / arbiter probes & 8 consumers / 30 kill points / 24+30+10 contention trials & Single-source location, durable trigger recovery, and single-holder arbitration were preserved under the corresponding perturbations \\
Guarantees & Frozen-surface split and required-role permanence & Released-package audit & 30 uninstall attempts + 3 wipes + 5 substitutions & No undeclared domain capability appeared in the APK; $L_1$ uninstall was refused; bootstrap recovery and same-function substitution were preserved \\
\addlinespace[2pt]
Ablation & Necessity of host-tracked rebinding and reversible cleanup & No rebinding / cleanup & 20 trials & Produced failures in 17/20 trials: 14 stale-binding and 3 residual-state; the remaining 3 trials showed neither. The stale/residual split reflects the trial mix, not mechanism strength. \\
Ablation & Necessity of write-ahead reconciliation & No write-ahead reconciliation & 30 kill--restart trials & Produced 12 reconciliation mismatches after kill--restart \\
Ablation & Necessity of required authorities & No required authorities & 1/2/4/8 consumers; 120+120 contention/race trials & Lost normalized single-source location semantics: the broker kept a single holder in 120/120 trials, whereas per-block acquisition split in 90/120, granted partial sets in 120/120, and let 480 foreign writes onto the bus \\
\bottomrule
\end{tabularx}
\end{table*}

Table~\ref{tab:eval-guarantees} should be read row-wise. The \emph{Guarantees}
rows correspond directly to the properties claimed in
\S\ref{sec:eval:guarantees}: safe assembly, mediated boundaries, coherent
dependency handling and replacement, transactional membership change,
single-source location and durable triggering, single-holder arbitration, and
the frozen-surface split. The \emph{Ablation}
rows then record the weakened variants used to support the necessity argument.
Taken together, these rows make explicit that the guarantees reported in the
main text are not merely descriptive properties of one implementation snapshot:
they are enforced under direct probes and fail in characteristic ways when the
corresponding mechanism is removed.

\begin{table*}[t]
\centering
\caption{Composition and openness ledger: benchmark outcomes, pipeline baselines, and robustness.}
\label{tab:eval-main}
\footnotesize
\setlength{\extrarowheight}{1pt}
\begin{tabularx}{\textwidth}{@{}>{\raggedright\arraybackslash}p{2.4cm}
                              >{\raggedright\arraybackslash}p{3.9cm}
                              >{\raggedright\arraybackslash}p{1.9cm}
                              >{\raggedright\arraybackslash}X@{}}
\toprule
Part & Item & Count / value & Result \\
\midrule
Composition & Benchmark cases & 50 & All admit concrete delivery oracles; 23 involve location/mobility; all 50 involve proactive triggering and at least one scarce channel/resource \\
Composition & FFH representability & 50/50 & All cases fit the five-dimensional composition space; no interpreter widening required \\
Composition & Benchmark-level comparison & FFH: 50 representable, 44 closed, 6 honest non-closures; Automation: 38 representable, 17 closed & FFH represented all benchmark cases and closed substantially more of them end to end; the automation baseline relied heavily on workarounds and left 12 cases outside its declared abstraction \\
Composition & Automation baseline & 15 direct / 23 workaround / 12 not representable & Only 17/50 closed end to end under the same delivery oracles; main deficits were log-derived absence checks, guarded multi-source joins, slot-aware delivery, asset-bound behavior, and multi-step or stateful fallback behavior \\
Composition & Substitution tests & 36 & FFH realized 30 admissible variants and refused 6 forbidden tuples; automation baseline was local in 3/36 only \\
\addlinespace[2pt]
Openness & Retrieval / commissioning / honest pre-termination & 17 / 27 / 6 & Retrieval closed 17; 27 entered commissioning; 6 terminated honestly before artifact production \\
Openness & Honest pre-terminations & 6 & Unsupported, unauthorable, and principled-refusal outcomes before artifact production \\
Openness & Tier routing & 4 / 22 / 1 & Tier~1 / Tier~2 / Tier~3 commissioned cases \\
Openness & Main-run closure & 44/50 & All 27 commissioned cases admitted and executed successfully \\
Openness & Pipeline baseline: retrieval-only & 17/50 & Closed only the reuse hits; all retrieval misses remained unresolved without commissioning \\
Openness & Pipeline baseline: author-without-preview & 4 affected cases & Would have admitted four machine-valid but semantically rejected artifacts that preview filtered out \\
Openness & Pipeline baseline: author-without-device-local-final-admission & 4 affected cases & Would have activated the same four artifacts without a final trust decision on the user's device \\
Openness & Preview rejections / revision burden & 4 / mean 1.4, median 1, max 3 & Four machine-valid artifacts were revised after preview rejection; benchmark includes retrieval-ambiguous, style-sensitive, and preview-dependent cases \\
Openness & Repeated-generation robustness & 46/50 & Ten commissioned cases re-run five times each; 92\% repeated-run closure, mean revisions 1.6 \\
Openness & Researcher restatements & 7/50 & One standardized restatement at most per case; no artifact editing, prompt patching, or override of validation/preview \\
\bottomrule
\end{tabularx}
\end{table*}

Table~\ref{tab:eval-main} should be read row-wise. The \emph{Composition} block
summarizes the benchmark-level evidence for composition coverage and
comparative expressiveness, complementing the main-text comparison of
Table~\ref{tab:eval-coverage} with the automation-baseline breakdown into
direct, workaround, and non-representable cases. The \emph{Openness} block
summarizes the retrieval-first openness pipeline, including retrieval,
commissioning, honest non-closure, the three pipeline baselines, preview
rejection, repeated-generation robustness, and bounded researcher intervention.

\paragraph{Architectural guarantees and necessity.}
It should be read as the appendix counterpart of
\S\ref{sec:eval:guarantees}. The first seven rows record direct enforcement
probes on FFH-full. These rows correspond to the contract and runtime
properties claimed in the main text: malformed artifacts are rejected before
activation; conforming controls are admitted; mediated over-reach is refused;
required-provider absence suspends dependents rather than crashing them;
replacement and unload preserve registry coherence; interrupted installs roll
back cleanly; and the required authorities preserve single-source location,
durable triggering, and single-holder arbitration under perturbation. The final
three rows record the weakened variants used for the necessity argument. Their
role is not to define alternative deployment targets, but to show that the
reported guarantees do not survive removal of the corresponding mechanism.

\paragraph{Automation baseline.}
The automation baseline is defined in Section~\ref{sec:eval} as the
recipe-style trigger-condition-action substrate against which FFH's
composition coverage and end-to-end closure are compared. As reflected in
Table~\ref{tab:eval-coverage} and Table~\ref{tab:eval-main}, the benchmark
includes cases that exercise retrieval ambiguity, multi-step coordinated
actions, durable state suppression, guarded multi-source joins, stateful
fallback behavior, and principled non-closure. The comparison should therefore
be read as a benchmark over a broader reactive-mobile class rather than as a
narrow reminder-only comparison.

\paragraph{Pipeline baselines.}
The pipeline baselines summarized in the \emph{Openness} rows of
Table~\ref{tab:eval-main} isolate the contribution of individual stages in the
retrieval-first openness path of \S\ref{sec:eval:openness}. The
\emph{retrieval-only} baseline disables commissioned authoring and therefore
tests how far reuse alone can close the benchmark. The
\emph{author-without-preview} baseline removes the semantic-fit gate provided by
user preview, exposing the cases in which machine-valid artifacts would still
have been unacceptable as rendered behavior. The
\emph{author-without-device-local-final-admission} baseline removes the final
device-local acceptance step, exposing the cases in which artifacts would have
become active without a final trust decision on the user's device. These
baselines are not intended as deployment targets; they are stage-removal
comparators showing that the openness pipeline's intermediate gates are
substantive, not ceremonial.

\paragraph{Openness protocol notes.}
The openness results of Section~\ref{sec:eval:openness} should be read with two
scope conditions in mind. First, researcher intervention was bounded: at most
one standardized restatement of the original intent was permitted per case, and
no artifact editing, prompt patching, or validation override was allowed.
Second, the evidence is strongest for Tier~1 and Tier~2 closure. Tier~3 is
represented by a single benchmark case and therefore supports a feasibility
claim for native-package introduction rather than a broad coverage claim over
the Tier~3 design space.

\paragraph{Cost measurement setup.}
The cost numbers of Section~\ref{sec:eval:tradeoff} are not repeated here;
the tables instead record the context in which they should be read. Cold start
and replay-based recovery were measured from process launch to reconstructed
steady state. Membership-change latency was measured for install and
replacement from activation to the manifest commit, and for unload from the
committed removal to disposal. Trigger-to-action latency was measured end to
end within the warm process (wake-path costs excluded) for periodic,
event-triggered, and geofence-triggered cases. Memory and
storage measurements on a debug build report 142\,MB PSS after startup (306\,MB RSS), a mean
per-plugin increase of 0.98\,MB, event-log growth of 0.23\,MB per 1{,}000
events, and 0.73\,MB of manifests and packaged rule assets for the shipped
assembly, excluding optional market packages. Medians are computed
over repeated runs, with p95 or a worst-case bound reported where repeated
samples exist.

\section{Capability-Level Comparison}
\label{app:comparison}

Table~\ref{tab:comparison} scopes the comparison claims of
Section~\ref{sec:related} to properties publicly demonstrated in the cited
sources. It is not a ranking of overall system quality. ``Not demonstrated''
denotes absence of public evidence for the listed property, not a claim of
impossibility.

\begin{table*}[t]
\centering
\caption{Capability-level comparison scoped to publicly demonstrated
properties in the cited sources.}
\label{tab:comparison}
\footnotesize
\setlength{\extrarowheight}{1pt}
\begin{tabularx}{\textwidth}{@{}>{\raggedright\arraybackslash}p{3.0cm}
                              >{\centering\arraybackslash}p{1.45cm}
                              >{\centering\arraybackslash}p{1.55cm}
                              >{\centering\arraybackslash}p{1.35cm}
                              >{\centering\arraybackslash}p{1.75cm}
                              >{\centering\arraybackslash}p{1.55cm}
                              >{\centering\arraybackslash}X@{}}
\toprule
System
& External plugin interface
& Runtime install
& Mobile-native
& End-user authorship
& Uniform contract
& Declared absence semantics \\
\midrule
FFH (this work)
& Yes
& Yes
& Yes
& Yes (Tiers~1--3)
& Yes
& Yes \\

DSH / Cordis \citep{shi2026composability}
& Yes
& Yes
& No
& Developer-oriented
& Yes
& Yes \\

Aura \citep{zou2026aura}
& Registry-mediated
& Yes
& Yes
& Not demonstrated
& Not demonstrated
& Not demonstrated \\

NLAH \citep{pan2026nlah}
& No
& Not demonstrated
& Not demonstrated
& Editable policies, not plugin authorship
& No
& Not demonstrated \\

AgentSkillOS \citep{li2026agentskillos}
& Curated skill package interface
& Yes
& Not demonstrated
& Public contribution, not end-user commissioning
& Not demonstrated
& Not demonstrated \\

Mobile-Agent-v2 / v3.5 \citep{wang2024mobileagentv2,xu2026mobileagentv35}
& No
& No
& Yes
& No
& No
& No \\

AppAgent \citep{zhang2023appagent}
& No
& No
& Yes
& No
& No
& No \\

MobiAgent \citep{zhang2025mobiagent}
& No
& No
& Yes
& No
& No
& No \\
\bottomrule
\end{tabularx}
\end{table*}

\end{document}